\documentclass[twocolumn,twoside]{article}
\usepackage{graphics}
\graphicspath{{Gamov2019}}
\usepackage[utf8]{inputenc}
\usepackage[english]{babel}
\usepackage{xspace}
\usepackage[colorlinks]{hyperref}
\def\xmm{XMM-Newton\xspace}

\begin{document}
\title{THE STUDY OF X-RAY SPECTRUM OF THE COMA GALAXY CLUSTER}
\author{L.Zadorozhna$^{1,2}$, A.Tugay$^{1}$, O.Prikhodko$^{1}$, D.Malyshev$^{3}$, Y.Sahai$^{1}$, D.Savchenko$^{4,5,6}$, N.Pulatova$^{7,8}$\\  
\date{
 $^1$ Faculty of Physics, Taras Shevchenko National University of Kyiv, Ukraine \\
 $^2$ Niels Bohr Institute, University of Copenhagen, Denmark\\
 $^3$ Institut f{\"u}r Astronomie und Astrophysik T{\"u}bingen, Universit{\"a}t T{\"u}bingen, Germany\\
 $^4$ Universit{\'e} Paris Cit{\'e}, CNRS, Astroparticule et Cosmologie, Paris, France\\
 $^5$ Bogolubov Institute for Theoretical Physics of the NAS of Ukraine, Kyiv, Ukraine\\
 $^6$ Kyiv Academic University, Kyiv, Ukraine\\
 $^7$ Main Astronomical Observatory, Kyiv, Ukraine\\
 $^8$ Max-Planck-Institute for Astronomy, Heidelberg, Germany\\
{\em e-mail: lidiia.zadorozhna@knu.ua, zadorozhna@nbi.ku.dk} }
}


\date{}
\maketitle
%
%
\begin{abstract}
The X-ray spectrum of the Coma galaxy cluster was studied using the data from the XMM-Newton observatory. We combined 7 observations performed with the MOS camera of \xmm in the $40'\times 40'$ region centered at the Coma cluster.
The analyzed observations were performed in 2000-2005 and have a total duration of $196$\,ksec.

We focus on the analysis of the MOS camera spectra due to their lower affection by strong instrumental line-like background.
The obtained spectrum was fitted with a model including contributions from the Solar system/Milky Way hot plasma and a power law X-ray background. The contribution of the instrumental background was modeled as a power law (not convolved with the effective area) and a number of Gaussian lines. The contribution from the Coma cluster was modeled with a single-temperature hot plasma emission. In addition, we searched for possible non-thermal radiation present in the vicinity of the center of the Coma cluster, originating e.g. from synchrotron emission of relativistic electrons on a turbulent magnetic field.
We compared the results with previous works by other authors and spectra obtained from other instruments that operate in the similar energy range of $1-10$\,keV. Careful and detailed spectrum analysis shall be a necessary contribution to our future work -- searching for axion-like particles’ manifestations in the Coma cluster.
\end{abstract}

$$~$$
%
%
{\bf 1. Introduction}\\[1mm]
Coma cluster (Abell 1656) is a hot gas galaxy cluster well-researched at all wavelengths. The Coma cluster's distance from the observer is $\sim 99$\,Mpc, redshift $z= 0.0231$, it contains above 1000 galaxies. It is virialized structure, with a mass about $10^{14} - 10^{15}M_{\odot}$ and spatial extents of $\mathcal{O}(1\,{\rm Mpc})$ ~\cite{Bower}. Apart from the dark matter component, which constitutes approximately 90\% of the cluster's mass, there is an additional approximately 10\% of the mass in a hot ionized intracluster medium (ICM). The intracluster medium accounts for the vast majority of a cluster's baryonic mass and produces diffuse X-ray emission due to thermal bremsstrahlung with a typical temperature of $T\sim 10^8$K and number density $n \sim 10^{-1}-10^{-3}$\,cm$^{-3}$ ~\cite{ChenY}.

Coma cluster's celestial size extends over more than $2^{\circ}$ on the sky. Its ICM is an extended X-ray source which size is above $45'$. As was evident from radio observations, the Coma cluster itself was formed through the merging of several smaller clusters. The central gas density $n_0=3.49 \cdot 10^{-3}$\,cm$^{-3}$ and virial parameters -- cluster's mass  $M_{vir}=1.2 \cdot 10^{15}M\odot$ and radius $R_{vir}=2.7$\,Mpc were obtained by {\L}okas and Mamon ~\cite{Lokas}.

Despite a plethora of observations, the characteristics and structure of the intergalactic plasma within clusters remain undefined. Both radio and X-ray observations indicate that the gas within these clusters is turbulent and disturbed ~\cite{Churazov1}, \cite{Schuecker}. The plasma within the clusters is weakly collisional ~\cite{Zhuravleva}, yet its collision rate is augmented by plasma instabilities. Furthermore, the transport processes are intricately tied to the local magnetic field's configuration. Multiple instances of X-ray and radio observations provide evidence for the existence of wave fronts, filaments, and bridges ~\cite{Churazov2021}. The substructures within the intercluster medium of Coma, notably the gas fluctuations caused by the collision of galaxy clusters, underscore the complex nature of the ICM.

Magnetic fields in ICM play a crucial role in cosmic ray emission and transport processes. The main ways to investigate a cluster's magnetic fields are the Faraday rotation measures (RM) and synchrotron radiation detection. Also, magnetic fields are interconnected with plasma and can persist in its internal structure. But at the same time, the properties of magnetic fields indicate their stochastic turbulent nature. Bonafede et al. estimated the magnetic field strength $B_0=4.7$\,$\mu$G from RM for the center of the Coma cluster ~\cite{Bonafede2010}. More recent radio observations from LOFAR and the Planck Observatory report correlations between radio synchrotron and soft X-rays, as well as radio -- Sunyaev-Zel'dovich (SZ) correlations, which may assist for better identification of the areas with a regular magnetic field ~\cite{BonafedeLOFAR}. 

The lowest limit of magnetic field strength was obtained by Wik et al. which is equal to $B = 0.2$\,$\mu$G ~\cite{Wik2009}. Generally accepted to use the beta-profile for gas density and for a profile of the magnetic field. The anisotropy parameter for Coma cluster $\beta= -0.75$ ~\cite{Lokas}, \cite{Bonafede2010} and $\eta=0.5$ ~\cite{Bonafede2}. The electron radial distribution $n_e(r) = n_{0}[1 + (r/r_c)^2]^{3\beta/2}$ and the magnetic field profile $B(r) = B_{0}[n_e(r)/n_0]^{\eta}$, where $n_{0}=3.44\cdot 10^{-3}$\,cm$^{-3}$, $r_c=291$\,kpc, $B_{0}=4.7$\,$\mu$G ~\cite{Bonafede2}.

For the last twenty years temperature map and a brightness profile of cluster core which is relatively homogenous ($8-10$\,keV) with a gradient from the northwest to the cooler southeast ($\sim 7$\,keV) have been obtained in ~\cite{Arnaud}, \cite{Gastadello}. 

Arnaud et al. obtained the X-ray spectrum of the central part ($\sim 0.78$\,Mpc) of the Coma cluster from \xmm EPIC/MOS (the European Photon Imaging Camera/Metal Oxide Semi-conductor) camera observations with total time 173\,ksec (5 overlapping observations)~\cite{Arnaud}. The best fit temperature is $kT = 8.25\pm0.1$\,keV and an abundance $0.25\pm0.02$, the absorption hydrogen column density $n_H = 9.4 \pm 0.9 \cdot 10^{19}$\,cm$^{-2}$. 

In the study conducted by Nevalainen and Lieu, the best-fit temperature was determined to be $kT = 9.2\pm0.7$\,keV, considering a systematic error of 5\% ~\cite{RLieu}. 

The existence of nonthermal emission in the X-ray spectrum of the Coma cluster was first supposed in the work of Fusco-Femiano et al.~\cite{FuscoFemiano} and this idea was developed in the articles of Nevalainen et al.~\cite{BeppoSAX} and Rephaeli et al.~\cite{YRephaeli}. Nevalainen et al. give the upper limit of nonthermal flux as 20 \% of the total flux and report the hard X-ray nonthermal electrons population in the energy range of 15-60\,keV ~\cite{BeppoSAX}. However, Wik et al.'s later research did not corroborate the previous findings in the hard X-ray range, failing to identify the expected Inverse Compton (IC) emissions~\cite{Wik2011}. Gastadello et al. determined the upper limit of the IC component within the Coma cluster's core ($12'\times12'$) at the energy range of $7-10$\,keV, resulting in a nonthermal flux value $5.1\cdot10^{-12}$\,erg\,cm$^{-2}$\,s$^{-1}$~\cite{Gastadello}. It's important to acknowledge that this value lacks precision due to the intense brightness of the thermal component at the cluster's center and the limited field of view (FoV).

Examination of the Rossi X-Ray Timing Explorer dataset from 1996 and 2000 validates that the utilization of thermal emission from isothermal gas does not yield a satisfactory fit for the spectral distribution of emissions within the inner $1^{\circ}$ radial zone Coma cluster. Although it's feasible to achieve spectrum alignment through emissions from gas exhibiting a notable temperature gradient, a more probable scenario encompasses the existence of an additional secondary nonthermal component. In such a scenario, it is estimated that nonthermal emission constitutes approximately 8\% of the entire $4-20$\,keV flux ~\cite{YRephaeli}.

Angus et al. provided a detailed review of the cluster soft X-ray excess in $0.2 - 0.4$\,keV and gave an alternative explanation of the phenomenon ~\cite{Angus}.

Consequently, the existence of nonthermal X-ray emissions at the cluster center remains subject to debate. However, the possibility of non-thermal emissions cannot be ruled out and necessitates further observations.

{\bf 2. Extraction and modelling of the spectrum} \\[1mm]
\xmm stands as a contemporary and advanced X-ray mission, currently in operation. It functions within an energy range of $0.2$\,keV to $12$\,keV and is equipped with two MOS and one PN cameras. These cameras exhibit a substantial total effective area that reaches its peak at around $2000$\,cm$^{2}$ at $1.5$\,keV, boasting a commendable energy resolution of roughly 10\% and a relatively extensive field of view spanning a radius of approximately $15'$ radius.

\begin{figure}[h!]
\resizebox{1.0 \hsize}{!}
{\includegraphics{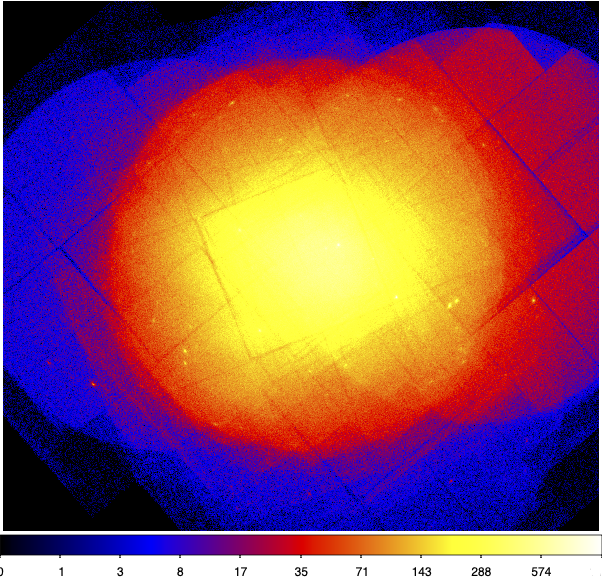}}
\caption{The picture of the Coma cluster is a mosaic of 8 partially overlapping pointings obtained with combined MOS1 and MOS2 cameras. 
Colors reflect the number of X-ray photons in a pixel.
}
\label{fig:coma_xmm}
\end{figure}

We analyzed publicly available observation data files for the Coma cluster from the XMM-Newton X-ray observatory. We used observations ObsIDs: 0124711401, 0153750101, 0300530101, 0300530301, 0300530401, 0300530501, 0300530601, 0300530701 with total exposure $343.8$\,ksec (see Fig.~\ref{fig:coma_xmm}). These data were processed utilizing the \href{https://www.cosmos.esa.int/web/xmm-newton/xmm-esas}{\emph{Extended Sources Analysis Software}} (ESAS) package, which is accessible as part of the Science Analysis System (SAS). Periods of time impacted by a notably variable background component, such as soft proton flares, were sieved using ESAS scripts named \texttt{mos-filter}.

We implemented the conventional filters and cuts criteria provided by the ESAS software suite. Notably, we eliminated prominent point sources identified through the standard SAS process \texttt{edetect\_chain}. To derive source spectra and create corresponding response matrices, we focused within the confines of a $14'$ radius circle centered around the source's position. These tasks were accomplished using the ESAS procedure \texttt{mos-spectra}. For the purpose of analysis, the obtained spectra were binned with an interval of $60$\,eV per energy bin, aiming to ensure that the bins remained statistically uncorrelated. We followed the procedure described in the article by Iakubovskyi et al~\cite{Iakubovskyi}. 

We modeled combined MOS spectra in \texttt{Xspec} spectral package. In our spectral analysis, we avoided the background subtraction procedure, since we are dealing with an extended object with a size exceeding the field of view and the proper selection of the region for the estimations of the background spectrum is not possible. We chose the modeled energy range $0.3-10.0$\,keV. To account for residual calibration uncertainties we added 1\% systematic error using \texttt{Xspec} parameter \texttt{systematic}. 

The resulting spectra consist of the Coma cluster's thermal radiation of plasma \texttt{apec\_3} and the astrophysical (Solar System plasma \texttt{apec\_1}, hot interstellar plasma \texttt{apec\_2}, cosmic X-ray background \texttt{powerlaw\_1}), the hydrogen column density for absorption \texttt{phabs} and the instrumental (smooth continuum \texttt{powerlaw} and line-like \texttt{gaussian} features) backgrounds. 

\begin{figure*}[h!]
\centering
\resizebox{1.2 \hsize}{!}
{\includegraphics{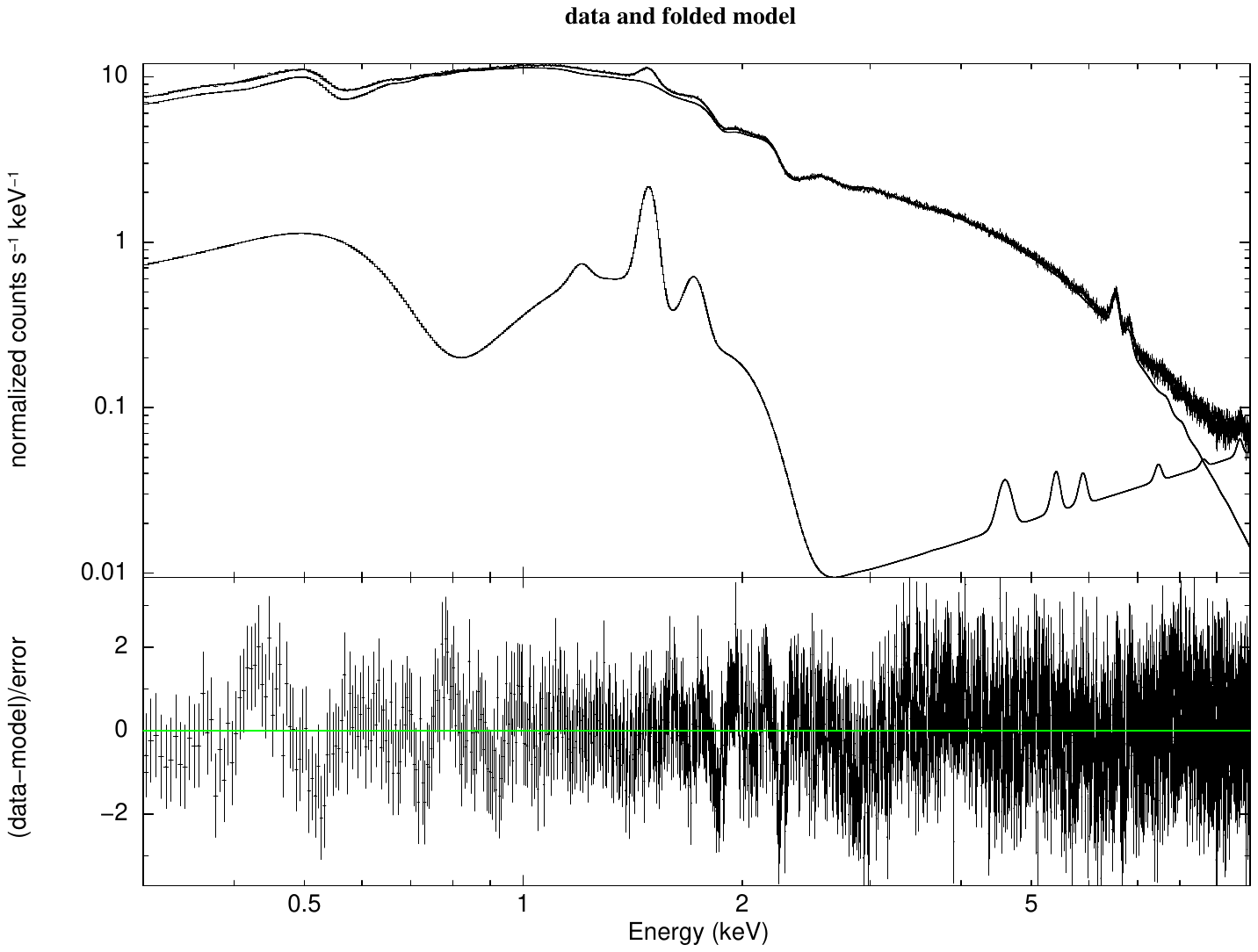}}
\caption{XMM-Newtom/MOS spectrum extracted from Coma cluster central region (top) and residuals with the best-fit model (bottom).
}
\label{xmm_spectrum}
\end{figure*}

We modeled the spectrum with the following \texttt{Xspec} syntax \texttt{apec\_1 + phabs*(powerlaw\_1+apec\_2+apec\_3)} presenting contributions from the described components. For instrumental background, we used \texttt{powerlaw+ gauss+...+gauss}, not convolved with the effective area. We included in the instrumental background model Gaussian lines, all were reported in \cite{Malyshev} (see Table~\ref{tab:xray_lines}). The fit parameters are summarized in Table~\ref{tab:best_fit} (Column {\bf Value fit 1}). The spectrum is shown in Figure~\ref{xmm_spectrum}.

A nonthermal correction was introduced to the aforementioned fitting models through the inclusion of a power law characterized by a freely adjustable spectral index parameter.
The normalization of this power law was set in a manner ensuring that the nonthermal contribution to the overall flux remained below 10\%. We modeled the spectrum using the following syntax \texttt{apec\_1 + phabs*(cflux\_1*powerlaw\_1 + apec\_2 + apec\_3+cflux\_2*powerlaw\_2)}. \texttt{cflux} is a convolution model to calculate the flux of other model components. The instrumental background was modeled similarly to the previous case. The summary of fit parameters can be found in Table 2 (Column {\bf Value fit 2}). An image of the spectrum was not presented due to the negligible alterations in comparison to the spectrum depicted in Figure~\ref{xmm_spectrum}.

\begin{table}
    \centering
    \begin{tabular}{ccc}
        \hline       
        Energy&Line&Origin\\
        \hline 
        0.56&O VII& astrophysical line\\
        0.65&O VIII& astrophysical line\\
        0.81&O VIII& astrophysical line\\
        0.91&Ne IX& astrophysical line\\
        1.34&Mg XI  & astrophysical line\\      
        1.49& 	Al-K$\alpha$& instrumental line\\
        1.56&    Al-K$\beta$&  instrumental line\\
        1.74&    Si-K$\alpha$&  instrumental line\\
        1.84& 	Si-K$\beta$& instrumental line\\
        2.11&    Au-M$\alpha$ &  instrumental line\\
        2.20&    Au-M$\beta$ &   instrumental line\\
        4.51& 	Ti-K$\alpha$& instrumental line\\
        5.41& 	Cr-K$\alpha$& instrumental line\\
        5.89&    Mn-K$\alpha$&  instrumental line\\
        5.95&    Cr-K$\beta$ &  instrumental line\\
        6.40& 	Fe-K$\alpha$& instrumental line\\
        6.49&    Mn-K$\beta$  &  instrumental line\\
        7.06&    Fe-K$\beta$  &  instrumental line\\
        7.48& 	Ni-K$\alpha$& instrumental line\\
        8.04& 	Cu-K$\alpha$& instrumental line\\
        8.26&  	Ni-K$\beta$  &  instrumental line\\
        8.63& 	Zn-K$\alpha$& instrumental line\\
        8.90& 	Cu-K$\beta$& instrumental line\\
        9.57& 	Zn-K$\beta$& instrumental line\\
        9.68&    Au-L$\alpha$&  instrumental line\\
                \hline       
    \end{tabular}
     \caption{The XMM-Newton background modeling incorporates the most salient instrumental and astrophysical lines. These lines have been adopted from the reference (TABLE I)~\cite{Malyshev}.}
    \label{tab:xray_lines}
\end{table}

{\bf 3. Conclusion}\\[1mm]
We merged a total of 8 overlapping observations captured by XMM/MOS within a $40'\times40'$ region around Coma cluster center.
For our analysis, we harnessed a cumulative observation time of 196 kiloseconds, extending across the years 2000 to 2005. The spectrum was fitted using two distinct models: a single-temperature thermal spectrum and a one-temperature thermal plasma model adjusted by means of a nonthermal correction. The inclusion of a nonthermal spectrum correction within $\sim 10$\% from total flux did not yield deterioration in the fitting's quality.

%
%
{\it Acknowledgements.} \\[1mm]
We are grateful for the discussions with Prof. Oleg Ruchayskiy and Prof. Bohdan Hnatyk. Lidiia Zadorozhna's work is funded by the Scholars At Risk Ukraine (SARU) fellowship at the University of Copenhagen. 

This research was made with the support of the Center for the Collective Use of Scientific Equipment "Laboratory of High Energy Physics and Astrophysics" of Taras Shevchenko National University of Kyiv. 
%
%
\renewcommand{\refname} 

\vfill
%

%
%

\begin{table*}
    \centering
 \begin{tabular}{clc|c|cccc}
  \hline
No & Model & Parameter & {\bf Value} & {\bf Value} \\ 
   & Component  & Unit &  {\bf fit 1} &  {\bf fit 2} \\ 
\hline
   1 & apec\_1 & kT,\,keV & $0.307\pm0.018$ & $0.303\pm0.020$ \\
   2 & apec\_1 & Abundance & 1.000 frozen & 1.000 frozen \\
   4 & apec\_1 & Redshift & 0.0 frozen & 0.0 frozen\\
   5 & apec\_1 & norm,\,cm$^{-3}$ & $(8.278\pm1.941)\cdot 10^{-4}$ & $(7.156\pm2.010)\cdot 10^{-4}$ \\
   6 & phabs   & $n_H$, $10^{22}$\,atoms\,cm$^{-2}$ & $(1.502\pm0.491)\cdot 10^{-2}$ & $(1.678\pm0.632)\cdot 10^{-2}$  \\
   7 & cflux\_1 & Emin,\,keV & -- & 0.200 frozen \\
   8 & cflux\_1 & Emax,\,keV & -- & 10.000 frozen \\
   9 & cflux\_1 & lg10Flux, flux in erg\,cm$^{-2}$\,s$^{-1}$ & -- & $-9.624\pm0.197$ \\  
   7 & powerlaw\_1 & PhoIndex & $1.478\pm0.021$ & $1.491\pm0.165$\\
   8 & powerlaw\_1 & norm,\,photons\,cm$^{-2}$\,s$^{-1}$\,sr$^{-1}$\,keV$^{-1}$ & $(2.450\pm0.249)\cdot 10^{-2}$ & 1.000 frozen \\
   9 & apec\_2 & kT,\,keV & $1.002\pm0.040$ & $0.998\pm0.041$ \\
  10 & apec\_2 & Abundance & 1.000 frozen & 1.000 frozen\\
  11 & apec\_2 & Redshift & 0.0 frozen & 0.0 frozen \\
  12 & apec\_2 & norm,\,cm$^{-3}$ & $(6.333\pm1.282)\cdot 10^{-4}$ & $(6.213\pm1.374)\cdot 10^{-4}$ \\
  13 & apec\_3 & kT,\,keV & $7.664\pm0.139$ & $7.656\pm0.140$\\ 
  14 & apec\_3 & Abundanc & $0.542\pm0.063$ & $0.628\pm0.252$\\
  15 & apec\_3 & Redshift &  0.0231 frozen &  0.0231 frozen \\
  16 & apec\_3 & norm,\,cm$^{-3}$ & $(7.658\pm0.908)\cdot 10^{-2}$ & $(6.628\pm2.686)\cdot 10^{-2}$\\ 
  7 & cflux\_2 & Emin,\,keV & -- & 0.200 frozen \\
   8 & cflux\_2 & Emax,\,keV & -- &10.000 frozen \\
   9 & cflux\_2 & lg10Flux,\,erg\,cm$^{-2}$\,s$^{-1}$ & -- & $<-10.4$ \\  
   17 & powerlaw\_2 & PhoIndex & -- & 2.000 frozen & \\
  18 & powerlaw\_2 & norm,\,photons\,cm$^{-2}$\,s$^{-1}$\,sr$^{-1}$\,keV$^{-1}$ & -- & $1\cdot 10^{-2}$ frozen \\
  
  \hline
       &     & $\chi^2$/d.o.f. & 1926.78/1895 & 1904.96/1908\\
       &     & null hypothesis probability & 30.0\% & 51.5\%\\
\hline
 \end{tabular}
  \caption{Model parameters of MOS1/MOS2 combined spectrum extracted from Coma cluster central region. Column {\bf Value fit 1} displays a set of parameters of the complex model \texttt{apec\_1 + phabs*(powerlaw\_1+apec\_2+apec\_3)} with adding the instrumental background. The column labeled {\bf Value fit 2} presents a collection of parameters from the intricate model with additional secondary nonthermal component \texttt{apec\_1 + phabs*(cflux\_1*powerlaw\_1 + apec\_2 + apec\_3+cflux\_2*powerlaw\_2)}, including the contribution of instrumental background. 
  Please note, that in absence of the detection of non-thermal (powerlaw) emission from the Coma cluster we present the limits on the flux of this component for the fixed to $2$ powerlaw index.}
  \label{tab:best_fit}
\end{table*}

\end{document}